\documentclass[11pt]{article}

\usepackage[utf8]{inputenc}
\usepackage[T1]{fontenc}
\usepackage[margin=1in]{geometry}
\usepackage{amsmath, amssymb, authblk} % authblk handles affiliations 
\usepackage{graphicx}
\usepackage[numbers,square,sort&compress]{natbib}
\usepackage[colorlinks=true, allcolors=blue]{hyperref}
\usepackage{lineno} % line numbers for Required for review
\usepackage{setspace} % For line spacing
\usepackage[font=small, labelfont=bf]{caption}%fontsize of caption

\title{\LARGE \textbf{Reconfigurable kirigami mesostructure enables modulation of lift and drag}}

\author[1,2, $\dagger$]{Agathe Schmider}
\author[3, $\dagger$]{Tom Marzin}
\author[1,*]{Sophie Ramananarivo}

\affil[1]{Laboratoire d’Hydrodynamique (LadHyX), CNRS, École polytechnique, Institut Polytechnique de Paris, 91120 Palaiseau, France}
\affil[2]{Unsteady Flow Diagnostics Laboratory, Institute of Mechanical Engineering, École Polytechnique Fédérale de Lausanne (EPFL), 1015 Lausanne, Switzerland}
\affil[3]{Department of Biological and Environmental Engineering, Cornell University, Ithaca, NY 14853, USA}
\affil[*]{e-mail: sophie.ramananarivo$@$polytechnique.edu}
\affil[$\dagger$]{A.S and T.M. contributed equally to this work}

\date{} % Leave blank as per Nature guidelines

\begin{document}

\maketitle
%\linenumbers % Start line numbering

% --- Abstract ---
\begin{abstract}
Flexible surfaces can modulate fluid forces through deformation, enabling passive adaptation to flow conditions. Here we show that kirigami sheets—planar surfaces patterned with arrays of parallel slits—provide a simple route to tunable aerodynamics by transforming into three-dimensional porous meso-architectures that can be reversibly reconfigured in flow. When exposed to cross-flow, parallel-cut kirigami buckle out of plane to form a lattice of inclined plate-like elements. Experiments reveal that this architecture generates not only drag but also a substantial transverse lift force, even when the sheet is held perpendicular to the incoming flow. Because the mesostructure can switch between distinct states, a single sheet produces large and selective variations in drag and lift under identical flow conditions, in some cases partially decoupling these forces. The evolving mesostructure also alters the scaling of forces with flow speed, influencing both instantaneous loads and their velocity dependence. Force measurements collapse when expressed in terms of the Cauchy number, identifying stiffness—set by the cutting pattern—as the dominant control parameter, a relationship captured by a continuum elastic model. These results show how kirigami architectures encode aerodynamic functionality and behavior directly through their structure, providing a scalable platform for surfaces with reprogrammable fluid forces.
\end{abstract}

% --- Main Body ---
\section*{Introduction}

Flexible surfaces are widely used in aerodynamic and hydrodynamic systems. Beyond reducing structural weight, compliance enables passive shape adaptation to flow fluctuations, providing greater versatility than rigid structures. In flow, flexible surfaces can reconfigure to reduce drag through streamlining and the reduction of frontal area, a strategy inspired by plant adaptation to strong flows \cite{vogel1989drag,alben2002drag,harder2004reconfiguration,schouveiler2006rolling,marzin2022shape}. Flexibility is also exploited in lift-generating structures such as sails and membrane wings in animal flight and micro-air vehicles \cite{thwaites1961aerodynamic,nielsen1963theory,song2008aeromechanics,tiomkin2021review,lian2003membrane}, where it can delay stall, adapt camber passively, and improve aerodynamic performance \cite{song2008aeromechanics,ramananarivo2011rather}. Most aerodynamic surfaces are nearly inextensible, yet extending this framework to highly deformable nonlinear materials may reveal new elastofluidic responses \cite{mathai2023shape}. Porosity also influences fluid forces in systems such as nets \cite{loland1991current,zhan2006analytical}, filters, fog collectors \cite{li2021aerodynamics,moncuquet2022collecting}, or even bristled insect wings and bird plumage \cite{muller1998air,kolomenskiy2020aerodynamic}. In most cases, pore morphology varies little under deformation, and aerodynamic effects are captured using modified effective force coefficients \cite{pezzulla2020deformation,jin2020distinct,kolomenskiy2020aerodynamic} or Darcy-type seepage models \cite{iosilevskii2011aerodynamics}. By contrast, systems in which deformation strongly alters pore morphology, thereby dynamically modulating permeability, remain less explored and offer opportunities for coupled fluid–structure responses \cite{gehrke2025coupled,louf2020bending,lamoureux2025kirigami}.

Kirigami, which introduces networks of cuts into thin sheets, provides a route to highly deformable porous structures, with a direct potential for fluid force modulation. The cuts enable large extensions through opening, while the mechanical response is governed by the cutting pattern, making kirigami a class of mechanical metamaterials with growing engineering interest \cite{callens2018flat,zhai2021mechanical,tao2023engineering,jin2024engineering}. Stretching generates porous architectures whose size and shape evolve with deformation, and that can extend into three dimensions through out-of-plane buckling \cite{isobe2016scirep,dias2017kirigami,rafsanjani2017buckling}. These three-dimensional textures have enabled applications ranging from directional friction to solar tracking and biomedical devices \cite{rafsanjani2018kirigami,babaee2020bioinspired,lamoureux2015dynamic,babaee2021kirigami}. In fluid-related contexts, such mesostructures have enhanced droplet capture in fog collectors \cite{li2021aerodynamics}, steered airflow for ventilation \cite{stein2024kirigami}, and created textured skins for aerodynamic control \cite{gamble2020multifunctional,wen2023dynamic}. In these examples, however, the porous texture is tuned externally and remains fixed under flow, rather than adapting to fluid loading.

Recent studies show that kirigami sheets in flow exhibit complex multiscale dynamics, where local interactions between the fluid and the evolving porous structure drive large-scale deformations \cite{marzin2022flow,carleton2024kirigami,lamoureux2025kirigami}. The inherent three-dimensionality of the post-buckling geometry enables asymmetric deformations of initially planar surfaces \cite{marzin2022flow}, suggesting new possibilities for fluid-force modulation. While early work demonstrated drag control for stabilizing the descent of kirigami-inspired parachutes \cite{lamoureux2025kirigami}, the asymmetric deformations emerging from specific cutting patterns suggest an untapped potential for lift generation. Furthermore, the multistability of certain kirigami architectures allows unit cells to switch between distinct deformation modes, creating a reconfigurable mesotexture \cite{yang2018multistable,khosravi2023phononic,janbaz2024diffusive}. By leveraging these local transitions to induce global shape variations, aerodynamic forces could be tuned intrinsically. This bypasses the need for macroscopic reorientation—such as varying the angle of attack—required by conventional aerodynamic systems like wings or sails.

\begin{figure*}[t!]
    \centering
    \includegraphics[width=\linewidth]{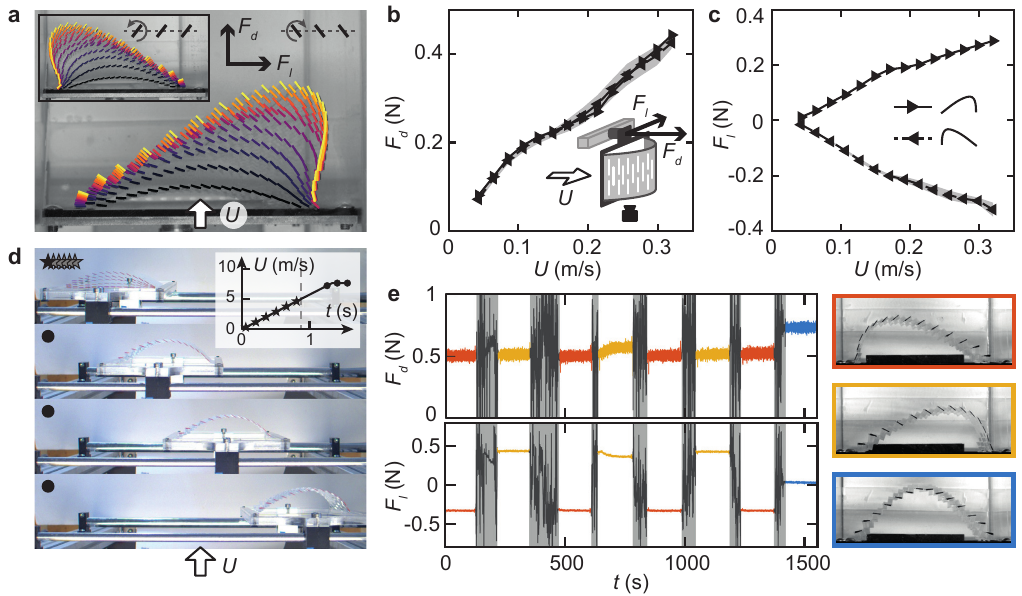}
    \caption{\textbf{Kirigami surface for tunable lift and drag generation.} \textbf{a} Overlaid images of kirigami deformation in a water flow at increasing speeds $U \in [4-32]\, \text{cm/s}$, with blades color-coded by velocity. Main: clockwise blade rotation; inset: counterclockwise for the same specimen. \textbf{b} and \textbf{c} Drag $F_d$ and lift $F_l$ as a function of flow velocity $U$. Right- and left-pointing triangles denote blade rotation direction. For the left-leaning case, data are averaged over 9 experiments, with shaded areas denoting the standard deviation. \textbf{d} As $U$ increases to 7.6 m/s (inset), a kirigami on a low-friction rail produces enough lift to move laterally in an airflow, with the dotted line in the inset indicating the onset of motion. \textbf{e} Real-time modulation of drag and lift in a water flow at fixed $U=23.5$ cm/s by manually switching blade direction: all clockwise (red), all counterclockwise (yellow) or opposite directions on each half (blue). Gray zones mark manual handling; data are averaged over a $4$ \textmu s sliding window.}
   \label{fig:1}
\end{figure*}

Here we investigate this concept using rectangular kirigami sheets patterned with staggered parallel slits. Stretching induces out-of-plane buckling of the uncut ligaments, forming a three-dimensional lattice of inclined plate-like elements. We show that this mesostructure allows a sheet held perpendicular to the flow to generate both drag and significant transverse lift. Combining experiments and theory, we map how these forces evolve with flow speed and how they depend on the cutting geometry. Importantly, the mesotexture can be reconfigured between distinct states for a fixed pattern, producing large and selective variations in drag and lift under identical flow conditions, as well as distinct scaling of forces with flow velocity. These results demonstrate that reconfigurable kirigami porous geometries provide a versatile platform for tunable and switchable fluid forces.

\section*{Results}

\subsection*{A lift and drag-generating surface}

\textbf{Experiments.} The kirigami specimen is fabricated by laser cutting a 100-\textmu m-thick Mylar sheet. The pattern consists of a periodic array of parallel slits (see Fig.\ref{fig:3}a), whose length and spacing are systematically varied to tune the resulting morphology and mechanical response, as discussed later. The values of the cutting parameters for all specimens tested in this study are provided in the SI. The sheet is clamped at both ends onto an aluminum frame and placed in a water tunnel with a $15\times15$ cm cross-section, producing a controlled uniform flow, with the specimen oriented perpendicular to the flow direction. The distance between the clamped extremities defines the specimen length ($L=12.5$ cm), and its height is $H=10.7$ cm. The frame supporting the kirigami is mounted on a six-component force sensor, allowing simultaneous measurement of the net drag in the flow direction and the lift in the transverse direction (see inset of Fig.\ref{fig:1}b, and "Methods" for details). The force contribution of the frame itself was measured separately and subtracted from the total signal. For each imposed flow velocity $U$ in $4-32$ cm/s, forces were recorded for 60 s at an acquisition rate of 1024 Hz, and the reported values correspond to time-averaged forces. The static deformation of the kirigami, which is invariant along its height, was simultaneously imaged using a camera positioned beneath the transparent section of the water tunnel (see "Methods"). The Reynolds number based on the characteristic length $L$, $Re_L=\rho UL/\eta$ (with $\rho$ and $\eta$ being the water density and dynamic viscosity, respectively), ranges from $5.10^3$ to $4.10^4$, placing the system within the inertial regime. Since the structure is inherently multiscale, another relevant Reynolds number can be defined using the width of an individual plate element $d_x$ (see Fig.\ref{fig:3}a), yielding $Re_{d_x} \sim 1.10^2-1.10^3$.

\medskip
\textbf{Drag and lift generation.} Figure \ref{fig:1}a illustrates kirigami deformation under increasing flow velocity. As the structure stretches, pores open and allow fluid to pass through, while uncut regions buckle into blade-like features (outlined in color), with the color scale indicating increasing flow speed. This reconfiguration leads to a clear departure from the quadratic drag scaling typical of rigid bodies in the inertial regime: in Fig.\ref{fig:1}b, the drag force $F_d$ instead exhibits a sublinear dependence on flow speed. This drag reduction originates from both the progressive pore opening and the alignment of the blades with the flow. We also note, in passing, a slight change in the drag trend around $U=17$ cm/s as the sheet's right end aligns with the flow (light purple profile), suggesting that kirigami shape morphing may enable subtle modulation of the drag trend. 

In the configuration shown in the main panel of Fig.\ref{fig:1}a, the blades rotate clockwise, causing the lobe to lean to the right. However, because the cutting pattern is left-right symmetric, plate elements can also buckle in the opposite direction, producing a mirror-image deformation (see inset of Fig.\ref{fig:1}a and \cite{marzin2022flow}). Both configurations produce nearly identical drag force evolutions (Fig.\ref{fig:1}b). To assess reproducibility, the experiment with the left-leaning configuration was repeated three times on three distinct specimens with identical cutting geometry. Between successive tests on the same specimen, it was flattened in an oven between two rigid plates to remove any residual plastic deformation. The force curves shown represent the mean over the nine measurements, with the shaded gray area indicating the standard deviation. The results demonstrate good reproducibility of the measurements.

Because the surface elements tilt relative to the flow, the fluid exerts forces not only in the streamwise direction but also transversely. As shown in Fig.\ref{fig:1}c, the kirigami sheet generates substantial lift forces, $F_l$, comparable in magnitude to the drag, despite initially being oriented perpendicular to the flow. As with drag, the evolution of lift with increasing velocity deviates from the quadratic scaling characteristic of rigid bodies, reflecting continuous shape reconfiguration, and a subtle change in trend is observed around $U=17$ cm/s. However, unlike drag, the direction of lift is highly sensitive to the orientation of the blades, with mirror-image deformations producing forces of equal magnitude but opposite sign.

We further test the lift generation in air. A kirigami sheet mounted on a low-friction rail is exposed to gradually increasing airflow in an open wind tunnel, with motion allowed only perpendicular to the flow (see Fig.\ref{fig:1}d, SI, Movie S1). As velocity increases, the kirigami surface inflates and drifts sideways, demonstrating substantial transverse lift forces despite being oriented normal to the incoming flow. Remarkably, simply switching the blade orientation reverses the motion, something a conventional sail can achieve only by changing its angle of incidence.

\medskip
\textbf{In-situ force tuning via mesostructure reconfiguration.} We demonstrated that the same kirigami sheet can produce lift forces of opposite sign, depending on the internal configuration of its mesostructure, namely the direction of blade rotation. We show that this mesostructure can be reconfigured \textit{in situ} to achieve real-time, reversible force switching at a fixed flow velocity ($U=$ 23.5 cm/s; see SI for details). Manually reversing the blade rotation (Movie S2) flips the lift from positive to negative, as shown in Fig.\ref{fig:1}e (yellow and red segments of the curve), while the drag remains essentially unchanged. The intermediate gray regions reflect high force fluctuations during manual handling. In the cases discussed so far, all blades rotate uniformly, but the rotation can also be varied along the sheet’s profile. In the final blue segment of Fig.\ref{fig:1}e, we reverse the rotation direction midway along the sheet. This produces zero net lift due to the resulting left–right symmetric deformation, while drag increases, as the blades orient more perpendicularly to the incoming flow. These observations confirm the potential of local blade rotation for active force modulation, a concept that will be explored further below. Next, we first investigate how geometric cutting parameters govern force generation.

\begin{figure*}[t!]
    \centering
    \includegraphics[width=\linewidth]{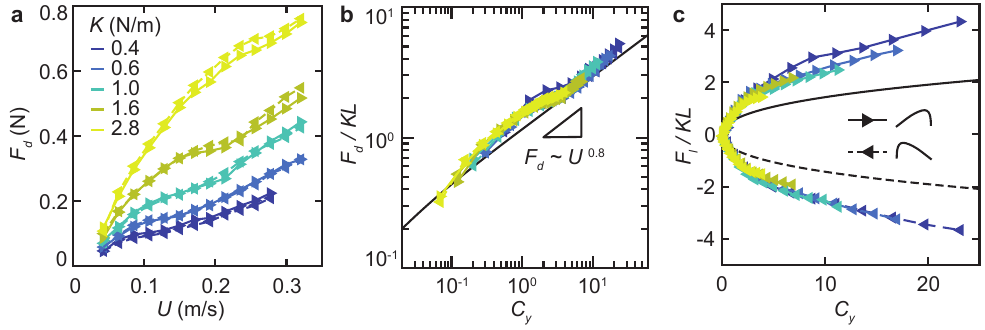}
    \caption{\textbf{Pattern-induced stiffness as a control parameter for aerodynamic performance.} \textbf{a} Drag as a function of flow velocity for kirigami specimens with different slit-row spacing $d_x$, yielding different effective stiffness $K$. \textbf{b} Dimensionless drag $F_d/KL$ as a function of the Cauchy number $C_y$ (Eq.\ref{Eq:Cy}), showing collapse and reduced velocity scaling compared to the quadratic law for rigid bodies. \textbf{c} Dimensionless lift $F_l/KL$ as a function of $C_y$; triangles indicate the blade rotation direction, producing opposite lift. In \textbf{b} and \textbf{c}, experimental results are compared with theoretical predictions (black lines).}
   \label{fig:2}
\end{figure*}

\subsection*{Effect of cutting parameters on forces through the effective sheet stiffness}

\begin{figure*}[t!]
    \centering
    \includegraphics[width=\linewidth]{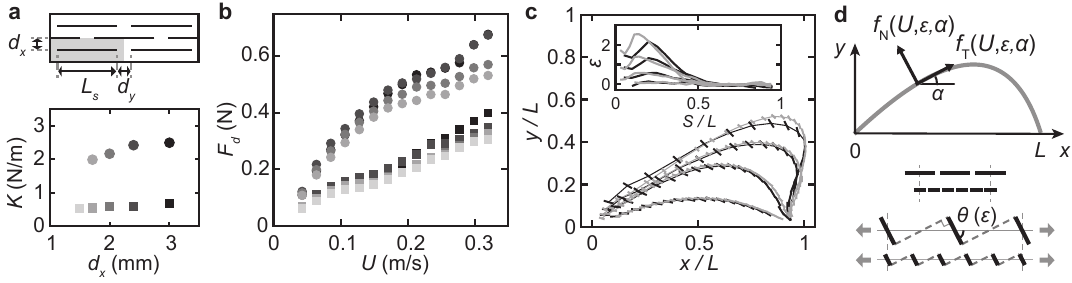}
    \caption{\textbf{Pattern independence at matched stiffness.}  \textbf{a} Stiffness from tensile tests for two iso-$K$ kirigami series (circle and square markers) with varying blade width $d_x$ (gray scale). \textbf{b} Similar drag evolution with flow speed within each series. \textbf{c} Profiles of two specimens with different blade widths, $d_x=1.7$ mm (gray) and $d_x=3$ mm (black), from the softer iso-$K$ series at velocities $U=4.3, 8.5, 14.9$ and $29.9$ cm/s. Inset: local elongation $\varepsilon$ along the curvilinear coordinate $S$ (defined in the unstrained flat configuration). \textbf{d} Membrane model with distributed normal and tangential fluid loading $\mathbf{f_{N}}$ and $\mathbf{f_{T}}$, which depend only on the flow velocity $U$ and on the local membrane state (elongation $\varepsilon$ and inclination $\alpha$). The forces $\mathbf{f_{N,T}}$ are set by the mesotructure (porosity and blade inclination $\theta$), determined solely by $\varepsilon$ and not by the cutting parameters, as illustrated schematically.}
   \label{fig:3}
\end{figure*}

\textbf{Effect of the cutting geometry.} The cutting pattern is defined by the slit length $L_s$ and by the spacing between rows and within rows of cuts, $d_x$ and $d_y$, respectively (Fig.\ref{fig:3}a). Changing the cut pattern notably modifies the mechanical properties of the sheet. Although kirigami sheets exhibit nonlinear responses, within the deformation range of our experiments, the tensile force varies linearly with extension and can be characterized by an effective stiffness $K$ \cite{marzin2022flow}. This stiffness can be related to the cutting parameters by modeling the kirigami sheet as an assembly of bending plates acting as springs arranged in series and parallel. The stiffness of these spring-like elements is primarily dictated by their geometrical dimensions, which are controlled by the cutting pattern, yielding a scaling for $K$ \cite{isobe2016scirep,hwang2018tunable,marzin2022flow}:
\begin{equation}
    K_{sc}=\frac{N_y}{2N_x}\frac{Et^3d_x}{(L_s-d_y)^3}
    \label{Eq:K}
\end{equation}
Here, $N_x$ and $N_y$ are the numbers of elementary cutting cells in each direction, determined by the sheet dimensions rather than being adjustable parameters (see SI). The Mylar sheet has a Young's modulus $E=4 \pm 0.3$ GPa and thickness $t=100$ \textmu m. To probe the influence of the cutting pattern on the fluid forces generated, we vary $d_x$ from $1.5$ to $4$ mm while keeping $d_y=5.7$ mm and $L_s=3$ cm constant. This results in stiffness values spanning $K=0.4- 2.8$ N/m, as measured by tensile tests (see "Methods").

The evolution of drag with flow velocity for specimens with different $d_x$ (i.e., different stiffness) is shown in Fig.\ref{fig:2}a. Both directions of blade rotation were tested, and as discussed previously, mirror-image deformations produce identical drag. As intuitively expected, stiffer specimens (larger $d_x$, yellow) generate higher drag than more flexible ones (smaller $d_x$, blue), which deform more and thus reduce drag through reconfiguration. More generally, the observed behavior reflects the balance between external fluid loading and internal elastic restoring forces. Following the approach of \cite{marzin2022flow}, we therefore recast the results in terms of the Cauchy number $C_y$, which quantifies this balance:
\begin{equation}
    C_y=\dfrac{\rho U^2H}{K}
    \label{Eq:Cy}
\end{equation}
When plotted against $C_y$, the dimensionless drag $F_d/KL$ collapses onto a single curve across diverse cutting geometries (Fig.\ref{fig:2}b), as does the dimensionless lift (Fig.\ref{fig:2}c), the latter changing sign for mirror-image profiles. These results indicate that the cutting pattern primarily acts through the effective stiffness it imparts, consistent with earlier observations for kirigami-inspired parachutes \cite{lamoureux2025kirigami}. Drag exhibits an approximate power-law scaling $F_d \propto C_y^{0.4}$ (i.e. $F_d \propto U^{0.8}$, obtained from a fit on all datasets), reflecting a substantially stronger drag reduction than that reported for kirigami parachutes ($U^{1.7}$). It falls between the regimes of bending plates ($U^{0.6}$) and the conical reconfiguration of a disk ($U^{0.9}$) \cite{gosselin2010drag,schouveiler2006rolling}. However, unlike these benchmark systems, which are anchored at a single point and bend along the flow, the present sheet is clamped at both ends. Consequently, drag reduction is driven by mesoscale blade reorientation rather than macroscopic streamlining, marking a distinct mechanism of fluid-structure interaction. 

Interestingly, the aerodynamic response appears to depend not on the details of the cutting geometry, i.e., the specific set of $(d_x,d_y,L_s)$, but solely on the resulting effective stiffness, a conclusion supported by the model introduced later. To test this hypothesis experimentally, we examine distinct cutting patterns designed to yield the same effective stiffness.

\medskip
\textbf{Insensitivity to pattern details at fixed stiffness.} We vary the blade width $d_x$ between 1.5 and 3 mm, adjusting $d_y$ and $L_s$ according to Eq.\ref{Eq:K} to maintain constant stiffness, thereby producing specimens labeled as "iso-$K$". Two iso-$K$ series with different target stiffness values were fabricated. Tensile tests reveal a slight increase in $K$ with $d_x$, especially for the stiffer series (Fig.\ref{fig:3}a), likely due to the simplifying assumptions of Eq.\ref{Eq:K}, such as neglecting edge effects. Indeed, non-uniform deformation near the boundaries affects the effective mechanical response of the stretched kirigami sheets \cite{taniyama2019design,taniyama2021design}. This effect becomes more pronounced when fewer unit cells span the sheet (i.e. larger $d_x$), as edge-affected regions then constitute a greater fraction of the surface.

Despite blade widths differing by up to a factor of two, each iso-$K$ series exhibits a similar evolution of drag with flow velocity (Fig.\ref{fig:3}b). Minor deviations are consistent with small variations in stiffness (see SI). As illustrated in Fig.\ref{fig:3}c, the corresponding deformations are likewise similar, both in terms of shape profile and local elongation (inset). Small differences arise from the edge effects mentioned earlier, with the specimen containing fewer unit cells (in black) exhibiting reduced elongation at the left end. This insensitivity of shape morphing and fluid forces to cutting details can be rationalized by a simplified model.

\medskip
\textbf{Theoretical model.} The behavior of the kirigami sheets in flow can be described using a reduced model that treats the structure as a continuous elasto-porous membrane, following the framework introduced in \cite{marzin2022flow}. The deformation is characterized by the local inclination angle $\alpha$ and the in-plane elongation $\varepsilon$ (Fig.\ref{fig:3}d). The equilibrium shape results from the balance between internal axial tension, assumed linearly elastic $N=KL\varepsilon$, and external fluid loading $\mathbf{f_{N,T}}$ acting along the normal and tangential directions of the profile, with unit vectors $\mathbf{n}$ and $\mathbf{t}$ \cite{marzin2022flow}:
\begin{equation}
  (N \mathbf{t})'+\mathbf{f_N}+\mathbf{f_T}=0
 % \label{Eq_mech}
\end{equation}
where $'$ denotes the derivative with respect to the curvilinear coordinate $S$ defined in the undeformed
state. Fluid loading is described by a semi-empirical expression, accounting for the mesotexture of tilted surface elements oriented at a local angle $\theta$ relative to the membrane
\cite{marzin2022flow} (Fig.\ref{fig:3}d):
\begin{equation}
  f_{N,T} = \tfrac{1}{2} \rho H C_{N,T}(\theta) \big(a(\varepsilon)\mathbf{U}.\mathbf{n}\big)^2
  \label{Eq_forces}
\end{equation}
Here, $\mathbf{U}.\mathbf{n}=U\cos \alpha$ is the normal velocity locally experienced by the membrane, and $a(\varepsilon)$ accounts for local blockage through flow rate conservation across the kirigami openings (see SI). The fluid-force coefficients follow semi-empirical relations  $C_N = C_{N0} \cos^3 \theta$ and $C_T= \text{sgn} \, C_{T0} \sin \theta \cos^2 \theta$, and $\text{sgn}=\{1,-1\}$ denotes the direction of blade rotation. The constants $C_{N0}=2$ and $C_{T0}=0.8$ were obtained experimentally \cite{marzin2022flow}. Importantly, the blade angle $\theta$ is geometrically determined by the local elongation through $\cos \theta = 1/(1+\varepsilon)$ \cite{lamoureux2015dynamic,marzin2023flow}, independently of the cutting parameters as illustrated in Fig.\ref{fig:3}d. Consequently, blade inclinations remain similar for the small- and large-blade specimens of Fig.\ref{fig:3}c, which share the same strain field. Local fluid forces in Eq.\ref{Eq_forces} therefore depend only on the elongation $\varepsilon$ (and the angle $\alpha$), such that the cutting geometry enters the model solely through the effective stiffness $K$, explaining the insensitivity to cutting details for patterns designed to have identical stiffness. 

Net drag and lift are obtained by integrating the local fluid stresses along the profile and projecting them onto the streamwise and transverse directions (see SI). The resulting predictions (black lines in Fig.\ref{fig:2}b and \ref{fig:2}c) agree reasonably well with experiments despite the model’s simplifications. Although forces are underestimated, which is consistent with the overprediction of sheet expansion reported in \cite{marzin2022flow}, the model captures both the observed trends and the velocity dependence. A fit over $C_y \in [0.2, 25]$ yields $F_d \propto C_y^{0.41}$, corresponding to $F_d \propto U^{0.82}$, in close agreement with the experimental scaling $F_d \propto U^{0.8}$.

\subsection*{Blade rotation for force modulation}

\begin{figure*}[t!]
    \centering
    \includegraphics[width=\linewidth]{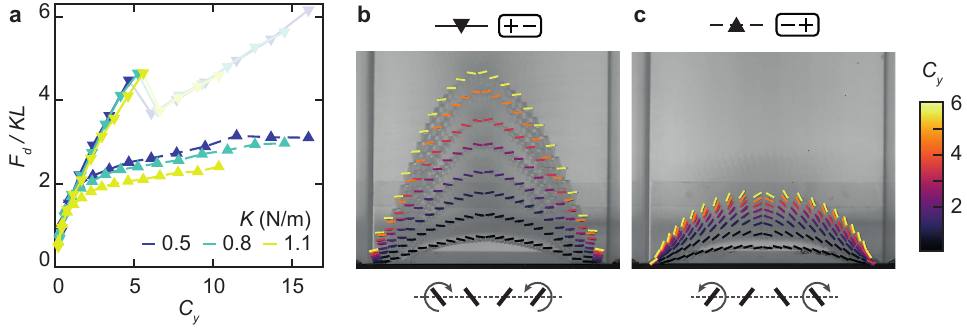}
    \caption{\textbf{Blade rotation as a mechanism for drag modulation.} \textbf{a} Dimensionless drag versus Cauchy number for three kirigami specimens of different stiffness (color-coded). Mid-sheet blade rotation reversal produces two configurations, ‘+ –’ and ‘– +’ (schematized in panel b; with ‘+’: clockwise, ‘–’: anticlockwise) indicated by upward and downward triangles, yielding distinct drag trends with velocity. \textbf{b} and \textbf{c} Symmetric profiles for the $K = 0.8$ N/m specimen under both rotation configurations.}
   \label{fig:4}
\end{figure*}

\medskip
\textbf{Influence of the mesostructure on force behavior.} 
In the previous section, all blades collectively buckle in the same direction, an energetically favorable state \cite{yang2018multistable,tang2017programmable}. However, as shown in Fig.\ref{fig:1}e, spatially modulating this rotation direction provides an effective method to tune the fluid forces. In this example, the mesostructure can maintain multiple stable configurations in flow and transition between them via external intervention. However, at low extension (either at low flow speed or in less-elongated regions) the energy barriers between stable states decrease, allowing adjacent blades to occasionally flip back to the same rotation direction. To prevent this and systematically study mesoarchitecture effects, blade orientation is first set using a small pre-test plastic deformation to maintain the desired rotational bias for the subsequent experiments. Secondary smaller cuts, similar to those in \cite{hwang2018tunable}, were also introduced to facilitate orientational transitions (see "Methods").

As an initial case, we reverse the blade rotation direction once, halfway along the kirigami. Figure \ref{fig:4}a shows the drag evolution for three specimens of different stiffness (color-coded), with markers indicating the direction of blade rotation in each half, '+ -' and '- +', where '+' denotes clockwise and '-' anticlockwise rotation. As before, drag data collapse when plotted in dimensionless form, confirming that the scaling with $C_y$ remains robust even for patterns with non-uniform blade rotation. Importantly, a marked difference in drag is observed between the two orientations. The '+ -' configuration (Fig.\ref{fig:4}b) produces an elongated lobe with blades nearly normal to the flow, generating drag up to 2.2 times larger than the '– +' case (Fig.\ref{fig:4}c), where blades align with the flow which strongly limits drag growth. By comparison, flipping a half-cylinder from convex to concave relative to the flow produces a similar change in drag \cite{hoerner1958fluid}. In contrast, here there is no overall inversion of the global geometry. Both configurations retain concave shapes; the difference lies primarily in the microstructure, highlighting the dominant role of local geometry in global force generation.

Beyond a critical Cauchy number, the ‘+ –’ configurations undergo a dynamic instability, manifesting as lateral oscillations that settle into a limit cycle. Reminiscent of previous observations of kirigami in airflows \cite{carleton2024kirigami}, our oscillations exhibit larger amplitudes and lower frequencies. These oscillations alter the drag evolution, as shown by the shaded points in Fig.\ref{fig:4}a which deviate from the steady-shape trend. Outside this unstable regime, static profiles remain symmetric, producing negligible net lift.

\begin{figure*}[t!]
    \centering
    \includegraphics[width=\linewidth]{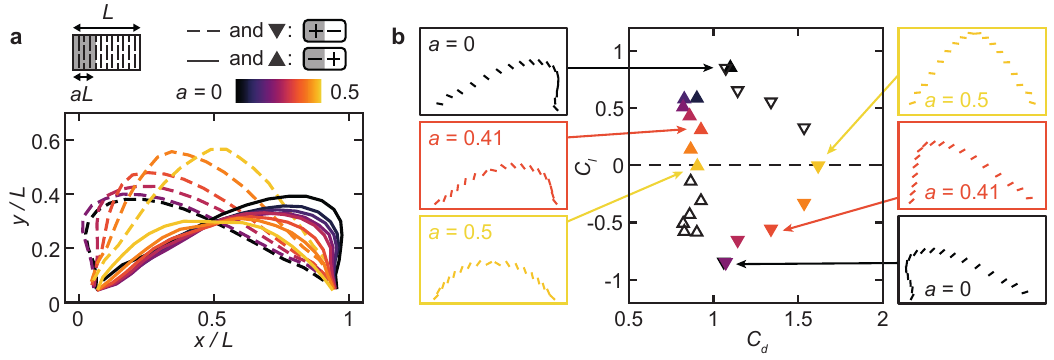}
    \caption{\textbf{Tuning of lift and drag via the mesostructural reconfiguration.} \textbf{a} Profiles for different positions of rotation reversal along the kirigami sheet, characterized by parameter $a$, for ‘+ –’ (dotted lines) and ‘– +’ (solid lines) configurations. \textbf{b} Lift versus drag coefficient for $a \in [0, 0.5]$ in both rotation configurations (upward and downward triangles), at $U=19.2$ cm/s. Values for the untested range $a\in]0.5, 1[$ are mirrored across the horizontal line (hollow triangles).}
   \label{fig:5}
\end{figure*}

\medskip
\textbf{Continuous tuning of lift and drag through the mesoarchitecture.} To explore the effect of the mesoarchitecture on fluid forces over a broader range of internal configurations, we systematically vary the blade-reversal location $aL$, with $a\in[0,1]$ (see schematics in Fig.\ref{fig:5}a). As the reversal point shifts from the center towards either extremity, the initially symmetric profiles (yellow in Fig.\ref{fig:5}a) gradually approach that of a uniform kirigami sheet (black), with an asymmetry set by the dominant rotation sign. We experimentally explored only half of the range, $a\in[0,0.5]$, since the remaining half produces mirror-image deformations. The corresponding fluid forces at a fixed flow speed $U=19.2$ cm/s are shown in Fig.\ref{fig:5}b, through the drag and lift coefficients: 
\begin{equation}
    C_{d,l}=\frac{F_{d,l}}{1/2\rho U^2HL}
\end{equation}
Because the experienced forces are not quadratic in flow velocity, $C_d$ and $C_l$ depend on $U$; the full evolution of forces with $U$ is provided in the SI. For the untested range $a\in]0.5,1[$, the inferred force values (empty triangles) correspond to mirror-image shapes, yielding identical drag but opposite lift. For uniform rotation ($a=0$, black triangles), we recover the same drag and opposite lift for left- and right-oriented lobes. For a central reversal ($a=0.5$, yellow triangles), symmetry cancels lift, while producing two distinct drag values. As the reversal location $a$ shifts, the data follow a closed trajectory in the $C_l-C_d$ plane, analogous to a polar curve in aerodynamics. In contrast to airfoils and sails, however, this evolution does not arise from a change in the global angle of attack (the kirigami mount is held fixed), but from microstructural reorientation through the parameter $a$. This closed lift-drag curve reveals a partial decoupling between lift and drag: distinct drag values can correspond to the same lift, and vice versa, solely through blade reorientation. Interestingly, for the '- +' orientation (upward triangles), as the profile evolves from a flattened shape to a one-sided lobe (solid lines in Fig.\ref{fig:5}a), the drag remains nearly unchanged while lift varies significantly. Such large lift variations at nearly constant drag are uncommon in classical aerodynamics, where both forces are typically strongly coupled through the global incidence.

\section*{Discussion}

Kirigami cutting techniques provide a simple yet versatile route to tune fluid forces in compliant surfaces. Rather than prescribing aerodynamic performance through global shape or incidence \cite{lasher2006interaction,lasher2008rans}, patterned cuts embed functionality directly within a reconfigurable material architecture. 

Two primary mechanisms underpin this behavior. First, the cuts significantly increase sheet compliance, shifting aerodynamic force scaling from the quadratic velocity dependence of rigid bodies to sublinear power laws. These trends collapse with the Cauchy number, which captures the balance between fluid loading and the elastic restoring forces engineered by the cuts. We find that the cutting pattern acts primarily through the effective stiffness it imparts: different geometries yield nearly identical aerodynamic responses when their stiffness is matched. A continuum elastic model, treating the kirigami sheet as a membrane under distributed fluid loading, captures both the force evolution with Cauchy number, and the insensitivity to pattern details at fixed stiffness, providing a simple analytical framework.

Second, cut openings generate a mesostructure of inclined plate elements through out-of-plane buckling. These elements produce substantial transverse lift, even with the sheet held perpendicular to the incoming stream, unlike most conventional porous materials. A kirigami “sail” can therefore move laterally in a tailwind. Importantly, this mesoarchitecture is reconfigurable and introduces a consequential design parameter: reversing the rotation of individual blades tunes lift and drag by more than a factor of two under identical flow conditions. This also enables a partial decoupling of lift and drag, whereby lift can be tuned with minimal drag variation, an uncommon feature in classical aerodynamics. While we manually trigger these transitions here, previous work has shown that they can be programmed using thermally responsive bilayers, enabling reversible switching and remote control \cite{tang2017programmable}. More complex spatial distributions of orientations could further expand the accessible force landscapes. Each row of elementary cells can switch between two rotation directions, effectively turning kirigami into an array of mechanical bits that encode a vast space of reprogrammable mesostructures \cite{khosravi2023phononic}.

Because blade orientation is intrinsically defined relative to the flow direction, reversing the flow alone would switch the force state. This contrasts with classical nets or porous sheets, whose drag response is typically symmetric upon flow reversal. A kirigami sheet placed in a channel could therefore function as a passive soft valve, generating asymmetric hydraulic resistance and enabling flow rectification or conversion of oscillatory forcing into net transport \cite{brandenbourger2020tunable}. Such drag asymmetry can also be exploited for propulsion, as appendages with direction-dependent drag produce net thrust under reciprocal actuation \cite{weathers2010hovering}. Kirigami surfaces thus encode macroscopic nonreciprocity in fluid–structure interactions directly through their microscale geometry.

Finally, microstructural tuning not only adjusts forces at fixed velocity but also reshapes their velocity scaling. Unlike other reconfigurable systems with fixed scaling exponents \cite{gosselin2010drag,schouveiler2006rolling}, a single kirigami sheet can modify its aerodynamic response curve through internal reorientation. Analogous to kirigami parachutes \cite{lamoureux2025kirigami}, these surfaces could regulate descent speed, while additionally enabling lateral steering through lift generation or mid-flight trajectory changes via microstructural reconfiguration.

Taken together, these results show that kirigami provides a versatile framework for encoding reprogrammable aerodynamic function directly into material architecture. The approach is lightweight, low-cost, scalable, and broadly compatible with soft materials, opening avenues for passive adaptive sails, smart nets, directional porous membranes, soft flow regulators, and deployable aerodynamic devices.

\section*{Methods}
\textbf{Experiments.} Experiments are performed in a closed-circuit water tunnel with a $15\times15$cm test section providing uniform flow. The channel has an open top for access and a removable ceiling to minimize free-surface effects, with a hole allowing connection of the kirigami to an external structure. Kirigami sheets are mounted on a rigid aluminum frame ($14.5\times14.5$ cm outer, $12.5\times12.5$ cm inner) attached to a six-component ATI Nano17 IP68 piezoelectric sensor, measuring drag and lift (range: 17 N in the flow direction, 12 N in the transverse directions; sensitivity: 1/320 N). The sensor is partially immersed to reduce the lever arm to the hydrodynamic center and prevent torque overloads, and mounted on a weighted structure atop an anti-vibration mat. After reaching steady-state deformations, forces are recorded for 60 s at 1024 Hz and averaged, with frame-only forces subtracted to isolate the kirigami response. Sensor drift, due to electrostatic discharge, is minimized by zeroing before testing each specimen and corrected by frame-only recordings, which exhibit the same drift. In longer experiments, such as real-time blade rotation reversal (Fig.\ref{fig:1}e), frame-only measurements show an approximately linear drift of $5.1. 10^{-3}$ N/min in the flow direction and $7.8.10^{-4}$ N/min tranversely, which was subtracted from the data. Kirigami deformation are imaged from below, using red tape on the blade tips for tracking. Blade positions and orientations are determined by fitting ellipses to contours detected via a custom Python-based image analysis routine (see SI).

\medskip
\textbf{Kirigami specimen and mechanical characterization.} Kirigami specimens are laser cut from 100-\textmu m-thick polyethylene terephthalate (PET) sheets using a Trotec Speedy 400 with a 120W CO$_2$ laser at 75\% power and 60\% speed. Specific cutting parameters for each experimental series are provided in the SI. To better control individual blade rotations, minor transverse secondary cuts are introduced at the tips of the longer slits (see SI for details). These cuts soften the hinge regions, relaxing slender plates from clamped-clamped to approximately pinned-pinned boundary conditions \citep{hwang2018tunable}, facilitating opposite rotations in adjacent blades. Final blade orientations are set via local plastic deformation at the hinges under stretching.

Effective stiffness is measured through uniaxial tensile tests using a Zwick \& Roell electromechanical machine with a 5 N load cell. Specimens are clamped in a sandwich fixture, with grips aligned to maintain slits horizontal and untwisted. Tests are conducted at a quasi-static displacement rate of 100 mm/min over three consecutive loading-unloading cycles. The first cycle differs slightly due to polymer shape-memory effects and is excluded from analysis. The resulting non-linear stress-strain response comprises three regimes: initial in-plane plate bending (high stiffness), intermediate out-of-plane buckling (low stiffness), and final hardening. Structures in flow primarily operate in the intermediate regime, where the linear force–displacement relationship defines the effective pattern stiffness, $K$, obtained via least-squares fitting over cycles 2 and 3. For specimens with blade reversal, mechanical tests are performed with blades preconfigured. While the minor cuts reduce $K$ compared with unmodified patterns of identical $d_x$, $d_y$, and $L_s$, $K$ follows Eq.~\ref{Eq:K} with a constant correction factor.

% --- Back Matter ---
\section*{Data Availability}
The data supporting the findings of this study are available within the paper and its supplementary information files.

%\section*{References}
\bibliographystyle{naturemag}
\bibliography{LIT} % Replace with your .bib file name

\section*{Acknowledgements}
We thank P. Hémon for helpful discussions on force measurement, J. Bicot and the Mecawet team at PMMH for providing access to their tensile test machine and water jet cutter, and V. Baslé (intern, LadhyX) for his contribution to the image processing code. We acknowledge support from a JCJC Agence Nationale de la Recherche grant (ANR-20-CE30-0009-01) to S.R. and support from the Agence Innovation Défense to T.M.

\section*{Author Contributions}
A.S., T.M. and S.R. designed research; A.S., T.M. and S.R. performed research; A.S. and T.M. contributed new analytic tools; A.S., T.M. and S.R. analyzed data; and A.S., T.M. and S.R. wrote the paper.

\section*{Competing Interests}
The authors declare declare no competing interest.

\end{document}